\documentclass{sig-alternate}
\usepackage{graphicx}
\usepackage{url}
\usepackage{balance}  

\begin{document}


\title{On the Feasibility and Implications of\\Self-Contained Search Engines in the Browser}

\numberofauthors{1}

\author{
\alignauthor
Jimmy Lin\\[1mm]
       \affaddr{University of Maryland, College Park}\\
       \email{jimmylin@umd.edu}
}

\maketitle

\begin{abstract}
JavaScript engines inside modern browsers are capable of running
sophisticated multi-player games, rendering impressive 3D scenes, and
supporting complex, interactive visualizations. Can this
processing power be harnessed for information retrieval? 
This paper explores the feasibility of building a
JavaScript search engine that runs completely self-contained
on the client side within
the browser---this includes building the inverted index, gathering
terms statistics for scoring, and performing query evaluation. The
design takes advantage of the IndexDB API, which is implemented by the
LevelDB key--value store inside Google's Chrome browser. Experiments
show that although the performance of the JavaScript prototype
falls far short of the open-source Lucene search engine, it is
sufficiently responsive for interactive applications.
This feasibility demonstration opens the door to interesting applications
in offline and private search across multiple platforms as well as
hybrid split-execution architectures whereby clients and servers
collaboratively perform query evaluation. One possible future scenario
is the rise of an online
search marketplace in which commercial search engine companies
and individual users participate as rational economic actors, balancing
privacy, resource usage, latency, and other factors based on 
customizable utility profiles.
\end{abstract}

\section{Introduction}

In nearly all deployments, search engines handle the vast bulk of
processing (e.g., document analysis, indexing, query evaluation) 
on the server side; the client is
mostly relegated to results rendering and interface
manipulations. This approach vastly under-utilizes the tremendous
processing capabilities of clients. For example, web browsers today
embed powerful JavaScript engines capable of running 
real-time collaborative tools~\cite{Gutwin:2011}, powering
online multi-player games~\cite{Chen:Xu:2011}, 
rendering impressive 3D scenes, supporting
complex, interactive visualizations,\footnote{\url{http://d3js.org/}}
enabling offline applications~\cite{Leblon:2010},
and even running first-person shooters.\footnote{\url{http://www.quakejs.com/}}
These applications take advantage of
HTML5 standards such as WebGL, WebSocket, and Indexed\-DB, and therefore do not require
additional plug-ins (unlike with Flash).

Can we apply this processing
power for information retrieval in interesting new ways? This
paper explores the feasibility of building a JavaScript search engine
that runs completely self-contained in the browser---this includes
parsing documents, building the inverted index, gathering terms statistics for scoring,
and performing query evaluation.

Is such a design merely a curiosity, or does it offer advantages over
traditional client--server architectures? Even as a
curiosity, this work explores how far browser technologies have
advanced in the previous decade or so, where they have emerged as a
viable platform for delivering rich user experiences. However,
browser-based search engines provide interesting
opportunities for information retrieval, both from
the perspective of enabling novel applications and opening up the
design space of search architectures. These possibilities are
detailed in Section~\ref{section:arch}.

In addition to discussing the implications of a browser-based
JavaScript search engine,
this paper describes the design and implementation of
JScene 
(pronounced ``jay-seen'', rhymes with Lucene),
an open-source proof-of-concept that illustrates the feasibility of these ideas.
JScene takes
advantage of the IndexedDB API, which is supported by a few modern
web browsers and implemented using LevelDB in Google's Chrome browser---the
result is a completely self-contained search engine that executes
entirely on the client side without any external dependencies.
The design of the prototype, detailed in Section~\ref{section:design}, highlights the
challenges of performing (relatively) large-scale data manipulations inside
the browser as well as the idiosyncrasies and limitations of JavaScript
for implementing standard information retrieval algorithms.
The focus of experiments in Section~\ref{section:experiments} is to evaluate the
feasibility of the idea in terms of index scalability and query latency.

As a reference, JScene is compared against the popular open-source Java
search engine Lucene; it should not be a surprise that JScene falls short of Lucene in
performance, but results nevertheless demonstrate that a pure JavaScript implementation is
sufficiently responsive to support interactive search capabilities.
The interesting applications and architectural possibilities enabled by
the in-browser concept suggests that such designs merit additional
exploration, especially since advances in browser-based technologies
will continue to narrow the performance gap between in-browser and native applications.

\section{Design Implications}
\label{section:arch}

Suppose it were possible to build an in-browser JavaScript search engine that
is fully self-contained and delivers reasonable performance:\ so what?
More than a technical curiosity, such a design promises to open up many
interesting possibilities in terms of applications and search architectures,
detailed below.

\subsection{Applications}

There are at least three advantages of self-contained, in-browser search engines that
enable interesting applications:

\smallskip \noindent {\bf Offline access.} One obvious advantage of
this design is that the user doesn't need to be connected to the
internet, so that documents are available for searching offline. One
can imagine a background process that continuously ingests web pages
that the user has visited in the recent past and updates an index of
these documents---search capabilities would then be available even if
the computer were disconnected from the network. Previous studies have
shown that a significant fraction of users' search behavior on the web
consists of ``refinding''~\cite{Tyler_WSDM2010}, or searching for
pages they had encountered before. Thus, a reasonably-sized
local index might achieve good coverage of many user
queries. While building and maintaining the index, there is no
reason why pages themselves can't be cached locally to provide direct
access to content offline. Note that such an application sidesteps
the typical issues associated with maintaining data consistency in a networked
environment:\ the pages are read only and users are already
accustomed to artifacts of page caching in modern search environments
(e.g., content divergence between live and cached copies).

Beyond ``vanilla'' web pages, it should be possible, via lightweight
connectors, to build integrated search capabilities over a multitude
of web-based services that are ubiquitous today. This would make it
possible to index content of email (from web-based email services),
online documents (e.g., Google Docs), calendar entries, contacts, to-do
lists, etc., providing what we call ``desktop search'' today
completely within the browser.

\smallskip \noindent {\bf Private search.} Another advantage of a
search engine that resides completely self-contained within the
browser is that there is no third party logging
queries, clicks, and other interactions. This
is particularly useful when a user has a collection of documents she
wishes to search privately---for example, when researching a medical
condition, some stigmatized activity, or other sensitive topics. This
scenario would be operationalized by coupling the in-browser search
engine with a focused crawler:\ the user would, for example, direct
the crawler at a collection of interest (e.g., a website with
medical information), and the search engine would then ingest documents
according to crawl settings. In practice, this might happen overnight while the
computer is otherwise idle, and the index would be ready for searching
the next day. An external party might still be able to infer search
intent based on the documents gathered, but the degree
of privacy can be controlled by the breadth of the crawl
(for example, crawling a site hosting medical information in its
entirety would hide the exact aliment). This
represents a time (and space) vs.\ privacy tradeoff that the user can
determine based on personal preferences.

\smallskip \noindent {\bf Multi-platform execution.} Although
the application scenarios described above could be
accomplished via native applications or browser plug-ins, the primary
advantage of a pure JavaScript implementation is the ubiquity of the
browser and the use of HTML5 standards. Although in reality the
situation is far more nuanced, the use of standards means that
applications are able to execute in any compliant browser. It is true
today that support for newer standards varies, but maturity in
terms of implementation will improve over time. Thus, a
browser-based application promises seamless execution across operating
systems and devices (laptops, tablets, mobile phones).

Building on HTML5 APIs also simplifies multi-device
synchronization. Since the search engine manages index
structures locally, providing a seamless experience as the
user moves from device to device requires a mechanism for
synchronizing data. One can imagine a cloud-based mechanism
for accomplishing this, which would also double as a 
backup service---this is fundamentally no different from
services today that synchronize bookmarks and other
browser-resident information across different devices.

One might argue that a cloud-based backend would defeat the privacy
advantage of the design, but it would be reasonably straightforward to
layer encryption on top of the synchronization and storage infrastructure.
The implementation would be similar to third party services today
that provide encryption for data stored in public clouds
(e.g., Amazon's S3).

\subsection{Architectures}

In addition to enabling new types of applications, the in-browser
search engine concept opens the door to a number of novel search
architectures:

\smallskip \noindent {\bf Load shedding.} From the perspective of a
commercial search engine company, which needs to continuously invest
billions in building datacenters, in-browser search capabilities are
appealing from the perspective of reducing server load. However,
``dispatching'' queries for local execution on the client's machine
may eliminate the
opportunity to generate revenue (i.e., via ad targeting), but this is
an optimization problem that search engine companies can solve. Although
the coverage of a local index would be miniscule compared to
the centralized index of a commercial search engine, for
particular classes of queries (such as the ``refinding'' queries
discussed above), the local collection might be
adequate (and it is possible that refinding queries provide fewer
ad targeting opportunities anyway).
The type of query, properties of the local index (summarized
perhaps via some content digest), current query load on the servers, network
latency, and even time of day may factor into the decision of whether
query evaluation is best performed on the server
or in the client's browser.

\smallskip \noindent {\bf Split execution.} Instead of purely
server-side or client-slide query execution, there are possibilities
for split execution where query evaluation is performed
cooperatively. One attractive possibility is search personalization,
where search results are specifically tailored to a user's interests
(e.g., consider the query ``Impala'' coming from a car enthusiast
vs.\ a database researcher). Architecturally, search personalization
might be conceived as a two step process:\ first, retrieve a set of
``generic'' results, and then reweight or rerank those results based
on user-specific features such as query and browsing
history~\cite{Bennett_2012}, social bookmarks~\cite{Noll_2007}, or
the user's ego-centric social network (in the case of social
search on Facebook, LinkedIn, Twitter, etc.)~\cite{Vieira_etal_CIKM2007,Curtiss_etal_VLDB2013}.
In this setup, the generic search results might be computed by a centralized
service, and the client would handle personalization. The types of
features and signals needed for personalization are exactly those that
could be gathered by an in-browser search engine. In fact, this design
has additional advantages in being able to leverage more intimate
aspects of a user's profile, features that the user would not be
comfortable sharing with a third party.

Another possible architecture could implement a recent conception of search as
progressive refinement, for example, the cascade
model~\cite{Wang_2011}. The general idea is that search proceeds in
stages, starting with ``cheap'' features (e.g., simple boolean term
matching) to progressively more ``expensive'' features (e.g.,
analyzing phrase relationships). The increased cost of each stage is
offset by considering a progressively smaller set of documents until
the final results are returned to the user. It might be possible to
offload the later stages of such a ranking model to the client,
where the joint optimization considers the size of intermediate results, how
expensive the features are, server load, and other factors.

\smallskip \noindent {\bf Distributed search marketplace.}
Synthesizing the ideas discussed above, one possible future scenario is the
emergence of a marketplace for hybrid models of centralized and distributed search
where optimization decisions are arrived at jointly by rational
economic actors driven by incentives.

For example, a commercial search engine company might offer an incentive
for a user to execute all or part of a search locally, in the form of
a micropayment or the promise of privacy (e.g., not storing the
queries and interactions). 
From the search engine company perspective, the value of
the incentive can be computed from the capital and operating costs of running datacenters,
revenue opportunities from ad targeting, etc. 
Commercial search engine companies have a clear sense of which searches
are ``money-makers'' (rich ad targeting opportunities) and which
aren't. Yet, all searches currently cost the companies money.
From the users' perspective, they can control in a fine-grained manner
their preferences for privacy and resource usage. 
The marketplace determines when the incentives on both ends align.
To the extent that ``money-loser'' queries overlap with the types of
queries that can be handled locally, such transactions are mutually beneficial.

This scenario describes a possible market-driven solution to issues of
privacy and concerns over data mining by internet services. Current attempts at
addressing these issues via legal and policy tools cannot escape
unintended consequences, and a market-based solution with explicit
incentives may be more effective. The key idea is for all parties involved
to clearly articulate their utility functions with respect to privacy,
latency requirements, coverage, resource usage, etc.\ and let market 
mechanisms determine the conditions under which transactions occur.
The expression of these preferences and tradeoffs could be as fine-grained
as individual queries or as coarse-grained as generic ``search profiles'',
depending on the context of the search.\footnote{\small Admittedly, this solution leaves aside
the issue of how to solicit preferences from lay users and help them understand
the implications of their decisions. However, consumer education is already a problem today,
so this proposed marketplace solution doesn't make anything ``worse''.}

\section{Feasibility Study}
\label{section:design}

Having discussed the interesting applications and architectural
possibilities of self-contained, in-browser search engines, it is now
time to address the practical question:\ Is such a design actually feasible?
Note that the goal of JScene, the proof-of-concept prototype described in this paper,
is to show that it is possible to build a pure JavaScript
in-browser search engine with {\it reasonable} performance---within
users' latency tolerance for interactive search. Of course, the
performance of the system will not come close to a custom-built search
engine (e.g., Lucene), but that's not the point;
a feasibility demonstration confirms that this general concept warrants
further exploration. To facilitate follow-on work, the JScene prototype
is released under an open-source license and available
to anyone interested.\footnote{\url{http://jscene.io/}}

At the storage layer, JScene depends on LevelDB,
an on-disk key--value store built on the same basic design
as the Bigtable tablet stack~\cite{ChangFay_etal_OSDI2006}. It is
implemented in C++ and was open-sourced by Google in 2011. The key--value store
provides the Chrome implementation of the Indexed Database (IndexedDB)
API, which is formally a W3C Candidate Recommendation (July
2013).\footnote{\url{http://www.w3.org/TR/IndexedDB/}} LevelDB supports
basic put, get, and delete operations on collections called ``stores''. Keys
are maintained in sorted order, and the API 
supports forward and backward iteration over
keys (i.e., range queries). 
Data are automatically compressed using the Snappy compression
library, which is optimized for speed as opposed to maximum
compression. The upshot is that inside every Chrome browser,
there is a modern key--value store accessible via JavaScript.
The JScene prototype takes advantage of LevelDB, as exposed via the IndexedDB API,
to store all index structures.

\subsection{Index Construction}

Nearly all keyword search engines rely on an inverted index, which
maps terms to postings lists. Each posting in a postings list
corresponds to a document that contains the relevant term and typically holds
other information such as the term frequency or term positions (to
enable phrase queries). The biggest challenge of an in-browser
JavaScript-based search engine is implementing the inverted index
using the provided APIs.

The IndexedDB API is built around key--value pairs, where values can
be complex JavaScript objects and keys can be JavaScript primitives, a
field inside the value object, or auto generated. In JScene, the postings
are held in a store called {\tt postings}, where the
key is a concatenation of the term and the docid containing the term,
and the value is the term frequency. For example, if the term
``hadoop'' were found twice in document 2842, the key would be
``hadoop+2842'' (with ``+'' as the delimiter) with a value of 2. In
the tweet search demo application (see Section~\ref{section:experiments}),
tweet ids can be used directly as docids, but in
the general case, docids can be sequentially assigned as
documents are ingested. A postings list
corresponds to a range of keys in the {\tt postings} store, and thus query
evaluation can be translated into range scans, which are
supported by IndexDB.

A few alternative designs were considered, but then rejected (at least for this prototype):\ it
seemed more natural to map each individual posting onto a key--value
pair as opposed to accumulating a list as the value of a single key
(the term), since in that case the indexer would need to rewrite the
value every time a term was encountered. Of course, it is possible to batch data and perform
term--document inversion in memory, but this adds considerable complexity that is perhaps
not necessary for a proof of concept. In many retrieval engines,
terms are mapped to unique integer ids, which allows the postings to
be more compactly encoded. Since with IndexDB the keys are strings, this
doesn't seem like much help.
Another design might be to store the hash value of the term,
thus creating uniform-length keys. However, this would require separately
storing the actual term in the value (to handle hash collisions), 
which would take up too much space.

Given this design, the indexer operation is fairly straightforward.
Each document is represented as a JSON object
and the entire collection is stored in an array. The indexer
processes each document in turn and generates key--value
pair insertions corresponding to the inverted index design described
above. All transactions in IndexDB are asynchronous, where the caller
supplies an {\tt onsuccess} callback function which is executed once the
transaction completes. Thus, a na\"{i}ve indexer implementation based
on a for loop that iterates over the documents would simply queue potentially millions of transactions,
completely overwhelming the underlying store. Instead, the indexer is
implemented using a chained callback pattern---the {\tt onsuccess} callback
function of a transaction to insert a key--value pair initiates the
next insertion, iterating through all tokens in a document and then
proceeding to the next document, until the entire collection has been processed.
This style of programming, although foreign in languages
such as C/C++ or Java, is common in JavaScript.

Separately, the index also needs to store document frequencies (the
number of documents that a term appears in) for scoring
purposes. These statistics are held in a separate store, aptly named
{\tt df}. The document frequencies are first computed by iterating
over the entire collection and keeping track of term statistics in a
JavaScript object (i.e., used essentially as a hash map). Once all
documents have been processed, all entries in the object are inserted
into the store, with the term as the key and the document frequency as
the value. Once again, the chained callback pattern described above is used for these
operations.

\subsection{Query Evaluation}

Once an inverted index is constructed, query evaluation algorithms
traverse postings in response to user queries to generate a top $k$
ranking of results. Query evaluation for keyword search, of course, is
a topic that has been extensively studied (see~\cite{Zobel_Moffat_2006}
for a survey). In the context of
this work, the goal is to explore the feasibility of in-browser query
evaluation using JavaScript, not raw performance per se. 
Thus, experiments in this paper used a simple approach based on
{\it tf-idf} scoring that requires the first query term to be
present in any result document. There are several reasons for this choice:\
First, previous work has shown that this scoring model works reasonably well in
practice~\cite{Asadi_Lin_TOIS2103}. Second, terms in a user's query
are often (implicitly) sorted by importance, and so it makes sense
to treat the first query term in a distinguished manner. Third, this
approach serves as a nice middle ground between pure conjunctive (AND)
and pure disjunctive (OR) query evaluation. Finally, this approach
lends itself to a very natural implementation in JavaScript described below.

The prototype query evaluation algorithm uses a very simple hash-based
approach in which a JavaScript object is used as an accumulator to
store current document scores, with the document id as the property
and the score as the value (essentially, a hash map). In the
initialization step, the document frequencies of
all query terms are first fetched from the {\tt df} store.
Next, a range query corresponding to the first query term is
executed, and all postings are scanned. The accumulator hash map is
initialized with scores of all documents that contain the term. After
the first query term is processed, the query evaluation algorithm
proceeds to the next query term, which results in another range scan;
for each posting, the accumulator structure is probed, and if the key
(document) is found, the value (document score) is updated. All query
terms are processed in this manner. At the end, the contents of the
accumulator are sorted by value to arrive at the top $k$.

Two details are worth discussing. First, this query evaluation
algorithm bears resemblance to the so-called SvS algorithm for postings intersection
(i.e., AND-ing of all query terms) that
cyclically intersects the next postings list with the current partial
results~\cite{Culpepper_Moffat_TOIS2010}. However, the standard
implementation takes advantage of binary search, skip lists, and other
techniques---given the limitations of the LevelDB API, 
it is not entirely clear how such optimizations can be
implemented in JavaScript. Second, the design of the IndexedDB API makes the
JScene query evaluation code somewhat convoluted. 
A range scan begins by acquiring a cursor, which is an asynchronous
operation with an associated {\tt onsuccess} callback. An object passed into
the callback provides a method that advances the cursor. Thus, the
entire query evaluation algorithm is implemented (somewhat awkwardly)
as chained callbacks:\ when the sequence of callbacks corresponding to
the processing of the first query term completes, it triggers the
range query for the second term and the series of callbacks associated
with that, and so on. As with indexing, this style of programming
is foreign to developers used to building data management systems
in C/C++ or Java.

\section{Experimental Setup}
\label{section:experiments}

Experiments used tweets from the recent Microblog evaluations at the
Text Retrieval Conferences (TRECs) sponsored by
the U.S.\ National Institute for Standards and Technology
(NIST). Specifically, the TREC 2011~\cite{mbOverview2011} and
2012~\cite{mbOverview2012} evaluations used the Tweets2011 collection,
which consists of approximately 16 million tweets gathered from
January 23, 2011 to February 7, 2011 (inclusive). Initial trials
showed that the collection in its entirety was too large for the current
implementation, so a smaller sub-collection was created via
random sampling (more details below). Of course, JScene is capable of
working with arbitrary text collections, although Twitter presents a
compelling application scenario.

Experiments were conducted on a 2012-generation Macbook Pro, with a
quad-core Intel Core i7 processor running at 2.7 GHz with 16 GB RAM and
a 750 GB SSD. The laptop contained all available upgrades at the time
it was purchased, and can be characterized as high-end
consumer-grade. The machine ran Mac OS X 10.9.2 with Google Chrome
version 33.0.1750.146.

As a point of reference, JScene was compared to the open-source
Java-based Lucene search engine
(version 4.7.0). To provide a fair comparison, the Lucene
queries were formulated to specify the same constraints as in JScene
(e.g., documents must contain the first query term), although Lucene
uses a document-at-a-time query evaluation algorithm that operates
differently. To ensure that both systems were processing the same
content, for JScene the collection was first tokenized with the Lucene
tools provided as a reference implementation in the TREC
evaluations\footnote{\url{http://twittertools.cc/}} and the resulting
tokens were then re-materialized as strings to create the JSON
documents. At indexing time, JScene simply split these
strings by whitespace without performing any additional processing,
which ensured consistent tokenization with Lucene.

Initial trials indicated that JScene would not be able to index the
entire Tweets2011 collection ($\sim$16 million tweets) within a
reasonable amount of time. Thus, for the experiments a smaller
collection comprising 1.12m tweets was created by random sampling. In
total, the documents contain 13.9m tokens with 1.74m unique terms,
occupying 140 MB on disk uncompressed.

Performance was assessed using
109 queries from the Microblog evaluations at TREC 2011 and 2012. These
queries represent information needs developed by NIST assessors, based
on their conception of what users might want to search for on
Twitter. A few examples are shown in Table~\ref{table:queries}. 
After stopword removal these queries average 2.9 terms, which is slightly
longer than typical web search queries.
In the actual TREC evaluations, the queries were associated with
timestamps indicating the query time; these were ignored and search
was conducted over the entire (sampled) collection.
Note that without actual
query logs from Twitter, it is impossible to test JScene on a
``realistic'' query load---but the TREC queries represent a
widely-accepted evaluation benchmark by the information retrieval
community. 

The relevant metrics in these experiments are indexing
and query evaluation speed. Evaluations did not include
effectiveness (e.g., precision) for a few reasons:\ the smaller
sampled collection means that there are fewer relevant documents,
which introduces noise in early precision measurements
(the typical effectiveness metrics for these types of tasks). Furthermore,
state-of-the-art ranking algorithms apply machine learning to a
candidate set of documents (e.g., from a basic {\it tf-idf} model); previous
experiments have shown that end-to-end effectiveness is relatively
insensitive to the quality of these intermediate results~\cite{Asadi_Lin_SIGIR2013}, which makes
an isolated component-level evaluation less meaningful.

\begin{table}
\begin{center}
\begin{tabular}{ll}
\hline
id & query \\
\hline
\hline
2 & 2022 FIFA soccer \\
36 & Moscow airport bombing \\
41 & Obama birth certificate \\
67 & Boston Celtics championship \\
86 & Joanna Yeates murder \\
\hline
\end{tabular}
\end{center}
\vspace{-0.5cm}
\caption{Sample queries.}
\label{table:queries}
\end{table}

\section{Results and Discussion}

It took JScene 644 minutes ($\sim$10.7 hours) to build the inverted index for 1.12m tweets
and another 152 minutes ($\sim$2.5 hours) to construct the document frequency table (both
averaged over two trials). As hinted before, this is about the largest
collection that can be reasonably indexed at once given the current
implementation, corresponding to the user leaving the laptop on overnight 
(although there are other
usage scenarios where the documents are indexed incrementally). 
While building the
inverted index, the Mac OS X Activity Monitor showed CPU usage oscillating roughly
between 15\% and 25\%, where the peaks correspond to LevelDB compaction events.
These utilization levels suggest that the process is IO bound (even though the
machine is equipped with an SSD). The LevelDB data for the postings
occupy approximately 1.6 GiB on disk, and the document frequency table
another 0.2 GiB.

Indexing results translate into a sustained write throughput of
around 360 postings per second. However, these figures are not directly comparable with
other performance evaluations of LevelDB because of at least two
reasons:\ first, it is unclear how much overhead JavaScript and the
IndexDB API introduce, and second, our chained callback
implementation means that the insertions were performed sequentially (i.e., synchronously),
which is known to be much slower than the standard asynchronous write
mode.\footnote{\url{http://leveldb.googlecode.com/svn/trunk/doc/benchmark.html}}

For reference, Lucene took 27 minutes to index the same
collection on a single thread (averaged over two trials). 
The on-disk index size is just 154 MiB. It is quite clear
that JScene indexing throughput falls far short of Lucene, and that
Snappy compression is far less effective than special-purpose
compression schemes designed specifically for search. This should not be surprising.

\begin{table}
\begin{center}
\begin{tabular}{lrrrr}
\hline
System & mean & median & \mbox{\hspace{0.3cm}}P90 & \mbox{\hspace{0.3cm}}max \\
\hline
\hline
JScene & 146 & 106 & 311 & 1058 \\
Lucene & 1.4 & 0.8 & 2.8 & 9.8  \\
\hline
\end{tabular}
\end{center}
\vspace{-0.5cm}
\caption{Query evaluation performance comparing JScene and Lucene; all values in milliseconds.}
\vspace{-0.1cm}
\label{table:latency}
\end{table}

Table~\ref{table:latency} compares query latency of JScene and Lucene
for the 109 queries from TREC 2011 and 2012:\ figures show mean,
median, 90$^{th}$-percentile, and max values (averaged over three trials).
Results for both are with a warm cache and Lucene ran in a single thread.
Note that in addition to per-query latencies
reported in the table, Lucene has a one-time start-up cost of around 9ms
for initializing index structures. In any realistic setup, of course,
this cost is amortized over many queries and thus inconsequential. In terms of the mean,
JScene is roughly two orders of magnitude slower than Lucene; the performance
gap is about the same based on the other metrics.
This is of course not
surprising since Lucene uses specialized data structures and has
received much attention from the open-source community in terms of
performance tuning. However, JScene is nevertheless reasonably responsive, with
query latencies within the range that users would expect for
interactive systems. Note that these experiments do not account for network
latencies that result from querying a remote service in a traditional
client--service architecture; factoring in network latencies
would narrow the performance gap between Lucene and JScene.

\begin{figure}[t]
\centering
\begin{center}
\includegraphics[width=1.0\linewidth]{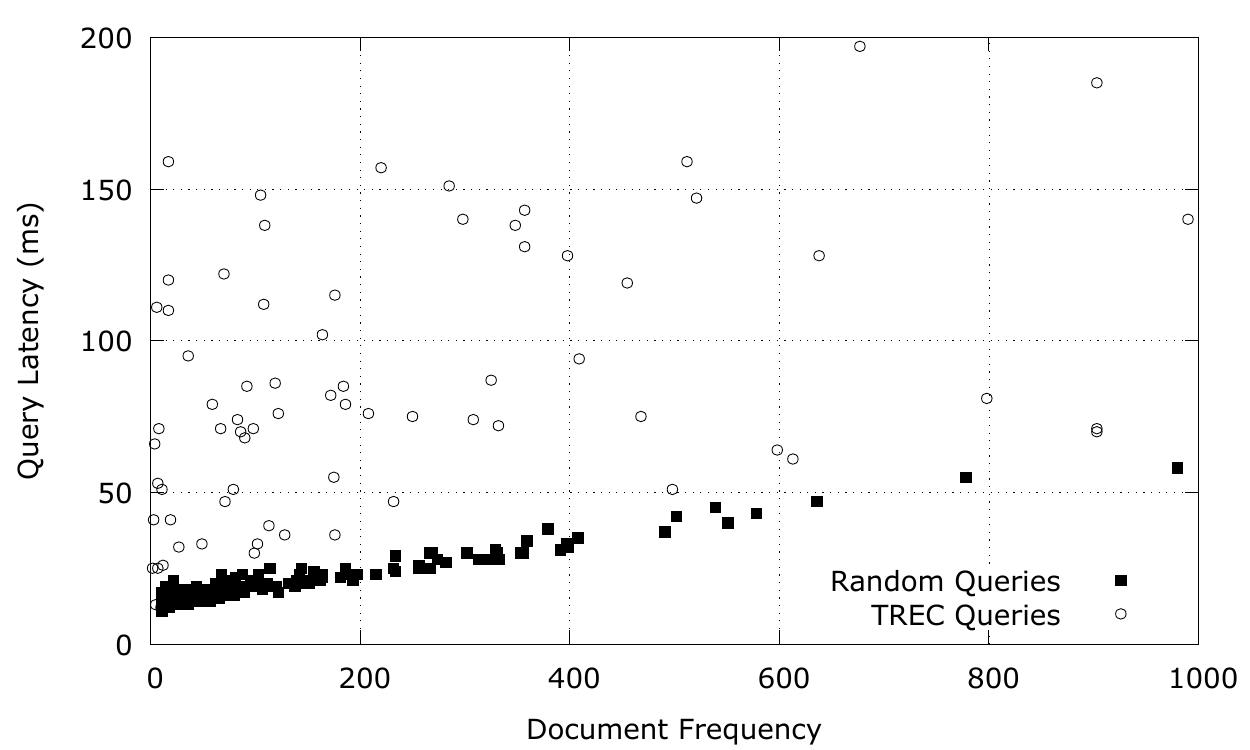}
\vspace{-0.5cm}
\caption{Latency vs.\ document frequency of query term for
  randomly-generated and TREC queries.}
\label{fig:performance}
\vspace{-0.3cm}
\end{center}
\end{figure}

To further explore the performance of JScene, a set of terms were
randomly sampled from the {\tt df} store and treated as single-term
queries. The performance of these queries are shown as solid squares
in Figure~\ref{fig:performance}. The figure focuses on terms with {\it
  df} less than 1000, but the linear relationship between {\it df} and
query latency extends to all sampled terms. The solid squares give a
sense of the lower bound of query latency, since any
document-at-a-time query evaluation algorithm will need to scan all
postings for a term. For comparison, the
TREC queries are plotted as circles based on the
{\it df} of their first query term. This plot illustrate two
points:\ First, there remains much room for improvement in JScene's
query evaluation algorithm. Second, and more importantly, even in the
most optimistic case, query evaluation with IndexedDB (via LevelDB),
at least with the current schema and storage layout,
will still be measured in tens of milliseconds.


\section{Future Work and Conclusion}

What can we conclude from these experiments? Results suggest that
although a self-contained, in-browser JavaScript search engine is much slower than a
custom native application (big surprise),
the JavaScript implementation is sufficiently
responsive for interactive querying.
The current prototype is sufficiently performant to be
deployed for searching (most) users' timelines, i.e., {\it all} tweets
that a user has ever read.
From this perspective, the design
is most definitely feasible and worthy of additional exploration.

These experimental results, however, reflect only a first attempt at
realizing the general concept. The prototype reflects a fairly
straightforward technical implementation, without applying any of the
standard efficiency ``tricks'' that are available in every
researcher's toolbox. These include various types of compression,
alternative schemas and storage layouts, optimizing data access
patterns for better locality, etc. These techniques, coupled with
future improvements in the IndexedDB implementation, will narrow the
gap between in-browser and native search applications.

In addition to query latency, index scalability (specifically,
throughput) is another performance concern with the present
prototype. However, keep in mind that the current limitations on
indexing speed apply only to batch indexing, where the system is
presented with the collection all at once. In many application
scenarios, document ingestion co-occurs with user activity, which has
a natural upper bound. Nevertheless, the optimizations
discussed above will also improve the scalability of the indexer.

Given this feasibility demonstration, it would not be premature to
start exploring some of the applications and architectures discussed
in Section~\ref{section:arch}. Performance and scalability will
continue to improve and become less and less of an issue---in the limit,
local indexes have an inherent performance advantage in eliminating
network latencies involved in communicating with remote hosts.
Perhaps in time, centralized commercial
search engines will no longer monopolize search-based access to
information in the way they do today, but co-exist with other
players in a search marketplace.

\bibliographystyle{abbrv}
\bibliography{JScene}

\end{document}